\documentstyle[12pt,epsf]{article}
\input epsf.tex

\begin{document}
\newcommand{\beq}{\begin{equation}}
\newcommand{\eeq}{\end{equation}}
\newcommand{\beqa}{\begin{eqnarray}}
\newcommand{\eeqa}{\end{eqnarray}}
\newcommand{\beqar}{\begin{eqnarray*}}
%--------------------------------------------------------
\hoffset = -1.4truecm \voffset = -1.5truecm
\date{}
\title{\bf Structure of Neutron Star  with a Quark Core}

\author{{\bf G.H. Bordbar \footnote{Corresponding author}\footnote{E-mail :
Bordbar@physics.susc.ac.ir}}, {\bf M. Bigdeli}  and {\bf T.
Yazdizadeh} \\
Department of Physics, Shiraz University,
Shiraz 71454, Iran\footnote{Permanent address}\\
and\\
Research Institute for Astronomy and Astrophysics of Maragha,\\
P.O. Box 55134-441, Maragha, Iran }

%%%%%%%%%%%%%%%%%%%%%%%%%%%%%%%%%%%%%%%%%%%%%%%%%%%%%%%%%%%%%%%%%%%%%%%%%

\maketitle

%%%%%%%%%%%%%%%%%%%%%%%%%%%%%%%%%%%%%%%%%%%%%%%%%%%%%%%%%%%%%%%%%%%%

\begin{abstract}
The equation of state of de-confined quark matter within the MIT
bag model is calculated. This equation of state is used to compute
the structure of a neutron star with quark core. It is found that
the limiting mass of the neutron star is affected considerably by
this modification of the equation of state. Calculations are
carried out for different choices of the bag constant.
\end{abstract}
%%%%%%%%%%%%%%%%%%%%%%%%%%%%%%%%%%%%%%%%%%%%%%%%%%%%%%%
\newpage
\section{Introduction}
Neutron stars (NS) are among the densest of massive objects in the
universe. They are ideal astrophysical laboratories for testing
theories of dense matter physics and provide connections among
nuclear physics, particle physics and astrophysics. The maximum
mass of a neutron star is a subject that several theoretical
astrophysicists have tried to compute it. Below a certain maximum
mass, degeneracy pressure prevents an object collapse into a black
hole. To calculate the maximum mass, we require enough information
about the composition of the star. Different compositions  lead to
different equations of state (EOS). When nuclear matter is
compressed to densities so high that the nucleon cores
substantially overlap, one expects the nucleons to merge and
undergo a phase transition to de-confined quark matter. Such a
system could be realized in two possible ways: (a) complete
strange quark matter stars (b) neutron stars with a core of quark
matter. Glendenning \cite{glen} has shown that a proper
construction of the nucleon-quark phase transition inside neutron
stars implies the coexistence of nucleon matter and quark matter
over a finite range of pressure. This has the effect that a core,
or a spherical shell, of a mixed quark-nucleon phase can exist
inside neutron stars.

In this work, we calculate the structure of the neutron star with
a quark core and also strange star, and  compare our results with
our pervious  works in which we investigated the NS structure
without a quark core \cite{bord}.
%-----------------------------------------------------------
\section{Quark Matter Equation of State}
For the de-confined quark phase, within the MIT bag model
\cite{chodos}, the total energy density is the sum of a
non-perturbative energy shift B (the bag constant) and the kinetic
energy for non-interacting massive quarks of flavors $f$ with mass
$m_{f}$ and Fermi momentum $k_{F}^{(f)}=(\pi
^{2}\rho_{f})^{\frac{1}{3}}$ where  $\rho_{f}$  is the quarks
density of flavor f:
\begin{eqnarray}
      \varepsilon_{Q}=\frac{3}{8}\frac{m^{4}c^{5}}{\pi ^{2}\hbar ^{3}}\left[
       x\sqrt{x^{2}+1}(2x^{2}+1)-\sinh ^{-1}x\right] +\frac{3\hbar
        c}{2\pi ^{2}}(\pi ^{2}\rho)^{\frac{4}{3}}+B,
        \label{eng3f}
 \end{eqnarray}
where $ x={\hbar k_{F}}/{mc}$ , $\rho_{s}=\rho_{d}=\rho_{u}=\rho$,
$\rho $ is baryon density, and  $\varepsilon_{Q}={E}/{V}$. We
assume in this work that u and d quarks are massless and the s
quark has a mass equal to $m=150 MeV$. The bag constant B, can be
interpreted as the difference between the energy densities of the
non-interacting quarks and interacting ones, which has a constant
value such as $B=55$ and $90 MeV$ in the initial model of MIT.
 Inclusion of perturbative interaction
among quarks introduces additional terms in the thermodynamic
potential \cite{fahri}. We try to determine a range of possible
values for B by using the experimental data obtained at the CERN
SPS \cite{heinz}.

According to the analysis of those experiments, the quark-hadron
transition takes place at about seven times normal nuclear matter
energy density ( $\varepsilon_{0} = 156MeVfm^{-3}$). We assume a
density dependent B. In the literature there are attempts to
understand the density dependence of B \cite{adami,jin,blasch};
however, currently the results are highly model dependent and no
definite picture has come out yet. Therefore, we attempt to
provide effective parameterizations for this density dependence.
Our parameterizations are constructed in such a way that at
asymptotic densities B has some finite value $B_{\infty }$:
\begin{eqnarray}
B(\rho)= B_{\infty }+(B_{0}- B_{\infty }) \exp \left[
-\beta(\frac{\rho}{\rho_{0}})^{2}\right]\cdot \label{bag}
\end{eqnarray}
The parameter $ B_{0} = B(\rho = 0)$ has constant which is assumed
to be $ B_{0} =400$ in this work. and $\beta$ is numerical
parameter usually equal to $\rho_{0}\approx 0.17fm^{-3}$, the
normal nuclear matter density. $B_{\infty }$ depends only on the
free parameter $ B_{0}$. In order to fix $B_{\infty }$, we proceed
in the following way:

Firstly, we use the equation of state (EOS) of asymmetric hadronic
matter characterized by a proton fraction $x_{p}=0.4$ and the
$UV_{14}+TNI$ potential. By assuming that the hadron-quark
transition takes place at the energy density $\varepsilon =
1100MeVfm^{-3}$, we find that hadronic matter baryon density is
$\rho_{t}=0.98fm^{-3}$(transition density) and  at values lower
than it the quark matter energy density is higher than that of
nuclear matter, while with increasing baryon density the two
energy densities become equal at this density and after that the
nuclear matter energy density remains always higher. Eq.
(\ref{eng3f}) for quark matter with two flavors u and d reduces
to:
 \beq \varepsilon _{Q}=\frac{3\hbar c}{4}\pi
^{\frac{2}{3}}\left[
\rho_{u}^{\frac{4}{3}}+\rho_{d}^{\frac{4}{3}}\right] +B,
\label{eng2f} \eeq
where $\rho_{d}=2\rho_{u}=2\rho$.

Secondly, we determine $B_{\infty }=8.99$ by putting quark energy
density and hadronic energy density equal to each other
$(\varepsilon_{Q}=\varepsilon|_{\rho=\rho_{t}})$.

Finally, we calculate the EOS for the three flavors quark matter
using
 \beq P=\rho
\frac{\partial \varepsilon_{Q} }{\partial \rho
}-\varepsilon_{Q}\cdot \label{eos3f}
 \eeq
%
%-------------------------------------------------------------------------
\section{Mixed Phase}\label{NLmatchingFFtex}
The hadron-quark phase transition takes place within a range of
baryon density values. In other words, the fraction of space
occupied by quark matter smoothly increases from zero to unity
when eventually the last nucleons dissolve into quarks. In this
case, we have a mixture of the hadron, quark and electron
background in the system. Glendenning's construction \cite{glen}
describes a global division of the baryon number between the two
phases. The equilibrium conditions, in the case where the geometry
of droplets is neglected, are those for bulk systems. The neutron
star matter is assumed to be â stable and charge neutral. Thermal
effects are not expected to play any important role in neutron
star cores. We neglect such effects and put the temperature $T =
0$. The equilibrium conditions for the quark matter droplet to
coexist with the nucleon medium are that pressure and chemical
potentials in both phases coincide. We choose pressure as an
independent variable. The coexistence requires that (Gibbs
conditions):
\beq \mu _{N}^{n}(P)=\mu _{N}^{q}(P), \eeq
and
\beq \mu _{P}^{n}(p)=\mu _{P}^{q}(P), \eeq
where $\mu _{N}^{n}$ and $\mu _{N}^{q}$  are the neutron chemical
potential in the nucleon phase (NP) and the quark phase (QP),
respectively. Similarly, $\mu _{P}^{n}$ and  $\mu _{P}^{q}$ are
the proton chemical potential in the respective phases.
The strange quark and lepton chemical potentials are dictated by
the conditions of weak equilibrium
\beq \mu _{s}=\mu _{d} \label{chsd}, \eeq
and
 \beq \mu _{d}-\mu_{u}=\mu _{l},\eeq
where $\mu _{l}$ is lepton chemical potential.

Using the semi-empirical mass formula, the energy per particle of
nuclear matter can be expressed as
 \beq
E(\rho ,x)=T(\rho ,x)+V_{0}(\rho )+(1-2x)^{2}V_{2}(\rho ),
\label{engnuc} \eeq
where $ x= {\rho_{p}}/{\rho}$ is proton fraction. The kinetic
energy contribution is
\begin{eqnarray} T(\rho ,x)=\frac{3}{5}\frac{\hbar ^{2}}{2m}\left( 3\pi
^{2}\rho \right) ^{\frac{2}{3}}\left[
(1-x)^{\frac{5}{3}}+x^{\frac{5}{3}}\right]\cdot \label{kennuc}
\end{eqnarray}
The functions $V_{0}$ and $V_{2}$ represent the interaction energy
contributions and we  can determine them from the results of
symmetric nuclear matter $(x=\frac {1}{2})$ and pure neutron
matter $(x=0)$ \cite{lagar}. Using our results for the nuclear
matter with the $UV_{14}+TNI$ potential, we have obtained the
following fit for $V_0$ and $V_2$:
\begin{eqnarray} V_{0}(\rho )=-559.9\rho ^{5}+1695.62312\rho
^{4}-1946.86437\rho ^{3}+1327.04\rho ^{2}-411.57428\rho -0.30327,
\end{eqnarray}
\begin{eqnarray} V_{2}(\rho )=191.21\rho ^{5}-626.712\rho ^{4}+776.623\rho
^{3}-473.40909\rho ^{2}+141.8709\rho +4.75638\cdot \end{eqnarray}
\\
From Eqs. (\ref{engnuc}) and (\ref{kennuc}), we obtain the
chemical potentials of neutrons and protons as:
 \begin{eqnarray} \mu _{N}^{n}(p)&=&\frac{\hbar ^{2}}{2m}\left( 3\pi
^{2}\rho \right) ^{\frac{2}{3}}\left[
(1-x)^{\frac{5}{3}}+x(1-x)^{\frac{2}{3}}\right] +V_{0}(\rho )+\rho
V_{0}^{^{\prime }}(\rho ) \nonumber \\&& \nonumber \\&&
+(1-2x)^{2}\rho V_{2}^{^{\prime }}(\rho )+(1-4x^{2})V_{2}(\rho
)+mc^{2}, \end{eqnarray} and
\begin{eqnarray} \mu _{P}^{n}(p)&=&\frac{\hbar ^{2}}{2m}\left( 3\pi ^{2}\rho
\right) ^{\frac{2}{3}}\left[
x^{\frac{5}{3}}+(1-x)x^{\frac{2}{3}}\right] +V_{0}(\rho )+\rho
V_{0}^{^{\prime }}(\rho ) \nonumber \\&& \nonumber \\&&
+(1-2x)^{2}\rho V_{2}^{^{\prime }}(\rho )+(-3+8x-4x^{2})V_{2}(\rho
)+mc^{2}\cdot \end{eqnarray}
The quark chemical potential with favor $f$ is
 \begin{eqnarray} \mu _{f}&=&\left[
m_{f}^{2}c^{4}+\hbar ^{2}c^{2}(\pi ^{2}\rho
_{f})^{\frac{2}{3}}\right] ^{\frac{1}{2}}\cdot
\label{muf}\end{eqnarray} From Eqs. (\ref{eng3f}), (\ref{eos3f}),
(\ref{chsd}) and (\ref{muf}), we obtain the chemical potential of
quark matter:
\begin{eqnarray}
\mu _{u}&=&\left( 4\pi ^{2}\hbar ^{3}c^{3}\left[ P+B+D-\rho
(\frac{\partial D}{\partial \rho }+\frac{\partial B}{\partial \rho
})\right] -\mu _{d}^{4}\right) ^{\frac{1}{4}}, \end{eqnarray}
where $D=\frac{3}{8}\frac{m^{4}c^{5}}{\pi ^{2}\hbar ^{3}}\left[
x\sqrt{x^{2}+1}(2x^{2}+1)-\sinh ^{-1}x\right]$. We also have
 \beq \mu
_{N}^{q}=2\mu _{d}+\mu _{u}, \eeq
and
 \beq \mu _{P}^{q}=2\mu _{u}+\mu _{d}\cdot \eeq
By plotting $\mu _{P}$ versus  $\mu _{N}$ for both nucleon and
quark phases, we can  identify the cross point of two curves that
satisfy the Gibbs condition. As the chemical potentials determine
the charge densities of the two phases, the volume fraction
occupied by quark matter, $\chi$, can be obtained by exploiting
the requirement of global charge neutrality:
 \beq \chi \rho _{q}^{c}+(1-\chi )\rho _{p}-\rho _{e}=0, \eeq
where $\rho _{q}^{c}$ is the quark charge density.
The total energy density and density of mixed phase (MP) are given
by:
\begin{eqnarray} \varepsilon
_{MP}=\chi \varepsilon _{QP}+(1-\chi )\varepsilon _{NP},
\end{eqnarray}
and
 \beq \rho _{MP}=\chi \rho _{QP}+(1-\chi )\rho _{NP}\cdot \eeq
We plot pressure versus baryon density for hadron, mixed and quark
phases in Figures 1 and 2, for bag constants $B=90$ and density
dependent $B$, respectively. It is seen that there is a mixed
phase at a range of densities. A pure quark phase is presented at
densities beyond this range and a pure hadronic phase is presented
at densities below it.
%----------------------------------------------------------------------------------
\section{Structure of Neutron Star  with a Quark Core}

We calculate the structure of a neutron star for various values of
the central mass density, $\varepsilon_{c}$, by using the equation
of state and numerically integrating the general relativistic
equation of hydrostatic equilibrium, Tolman-Oppenheimer-Volkoff
(TOV) equation\cite{shap}. The derivation of TOV-equation can be
found in standard textbooks \cite{nglen,weber,adler,misner}.

For a neutron star with a quark core, we use the following
equations of state:
\begin{itemize}
\item Below the density $0.05 fm^{-3}$, we use the equation of
state calculated by Baym et al. \cite{baym}.

\item From this density up to the beginning point of the mixed
phase, we use the equation of state which is calculated with the
$UV_{14}+TNI$ potential \cite{bord}.

\item For the range of densities in the mixed phase, we use the
equation of state  which was calculated in the previous section.

\item Beyond the end point of the mixed phase, we use the equation
of state of pure quark matter which was calculated in section 2.
\end{itemize}

Calculations are done both for constant $B=90$, and density
dependent $B$.  Based on these EOS's, we  calculate the mass and
radius of the NS with quark core. The calculations are also
repeated for the strange star (i.e. pure quark matter). We plot
the NS mass versus central mass energy density for B=90 and
density-dependent B, in Figures 3 and 4, respectively. The NS mass
versus radius for quark core NS and strange star are plotted for
B=90 and density-dependent B in Figures 5 and 6, respectively. For
the sake of comparison, we have also plotted our previous results
of the neutron star structure without quark core, in these
figures. It is seen that there is a profound difference between
the new results for NS with a quark core and those of NS without a
quark core.

The extracted maximum mass of a NS and the corresponding radius
and central mass density for both cases B=90 and density dependent
B are presented in Tables 1 and 2, respectively. It is seen that
the inclusion of the quark core leads to a considerable reduction
of the maximum mass, while the radius is not affected appreciably.
Note that the maximum mass for the NS with quark core and B=90 is
quite near to the observed maximum mass of neutron stars
\cite{Thorsett99}.

The maximum mass energy density versus the radial coordinate for
NS without core, NS with a quark core and strange star are plotted
in Figures 7 and 8 for B=90 and density-dependent B, respectively.
It can be seen that a major part of the core is composed of pure
quark matter (about 8 Km). A layer of mixed phase (thickness about
1.5 Km) exists between the core and a thin crust.

\section{Summary}
As we go from the center toward the surface of a neutron star, the
state of baryonic matter changes from the de-confined quark-gluon
to a mixed state of quark matter and hadronic matter, and thin
crust of hadronic matter. The transition between these states
occurs in a smooth way. In order to calculate the structure and
the mass limit of neutron stars, it is important to have a fairly
accurate physical description of these states.

In this paper, we calculated the equation of state of the
de-confined quark phase within the MIT bag model. We then
calculated the mixed phase of nucleons and quarks. The equilibrium
volume fractions of nucleon and quark matter were obtained by
applying the Gibbs condition. Curves were presented which showed
the dependence of pressure on the baryon density.

Our results for the equation of state were then used to calculate
the structure of a neutron star with a quark core. As usual, the
Tolman-Oppenheimer-Volkoff equation were integrated from the
center to the surface of the neutron star where the density drops
to zero. Calculations were carried out both for B=90 and a
density-dependent B.

The maximum mass, radius, and central mass density of neutron
stars with a quark core and strange star were calculated and
compared with the traditional neutron star. It was found that the
limiting mass decreases when the quark core is taken into account.
This brings the maximum mass closer to the observational limits.
\section*{Acknowledgements}
This work has been supported by Research Institute for Astronomy
and Astrophysics of Maragha, and Shiraz University Research
Council.

%%%%%%%%%%%%%%%%%%%%%%%%%%%%%%%%%%%%%%%%%%%%%%%%%%%%%%%%%%%%%%
\newpage
%%%%%%%%%%%%%%%%%%%%%%%%%%%%%%%%%%%%%%%%%%%%%%%%%%%%%%%%%%%%%%%%%%%%
%%%%%%%%%%%%%%%%%%%%%%%%%%%%%%%%%%%%%%%%%%%%%%%%%%%%%%%%%%%%%%%%%%%%

%%%%%%%%%%%%%%%%%%%%%%%%%%%%%%%%%%%%%%%%%%%%%%%%%%%%%%%%%%%%%%%%%%%%%%%%%%%%%%%%%%%%%%%%%%%%%%%%%%%%%%%
\newpage
\begin{table}
  \centering
  \caption{Maximum gravitational mass $(M_{\max })$, corresponding radius(R)
  and central mass density$(\varepsilon_{c})$ for $B=90$.}\label{1}
  \begin{tabular}{cccc}
  \hline
  star & $M_{\max }(M_{\odot})$ & R(Km) & $\varepsilon_{c}(10^{14}gr/cm^3)$ \\
  \hline
  NS & 1.98 & 9.81 & 27.17 \\
  NS+quark core& 1.57 & 9.73 & 33.27 \\
  stange star & 1.34 & 7.77 & 34.81 \\
  \hline
\end{tabular}
\end{table}

%%%%%%%%%%%%%%%%%%%%%%%%%%%%%%%%%%%%%%%%%%%%%%%%%%%%%%%%%%%%%%%%%%%%%%%%%%%%%%%%%%%%%%%%%%%%%%%%%%%%%%%%%%%
\begin{table}
  \centering
 \caption{Maximum gravitational mass $(M_{\max })$, corresponding
 radius(R)and  central mass density$(\varepsilon_{c})$ for density dependent $B$.}\label{2}
  \begin{tabular}{cccc}
  \hline
  star & $M_{\max }(M_{\odot})$ & R(Km) & $\varepsilon_{c}(10^{14}gr/cm^3)$ \\
  \hline
  NS & 1.98 & 9.81 & 27.17 \\
  NS+quark core& 1.75 & 9.66 & 28.92 \\
  stange star & 1.63 & 8.2 & 28.92 \\
  \hline
\end{tabular}
\end{table}
%%%%%%%%%%%%%%%%%%%%%%%%%%%%%%%%%%%%%%%%%%%%%%%%%%%%%%%%%%%%%%%%%%%%%%%%%%%%%%%%%%%%%%%%%%
\newpage
%%%%%%%%%%%%%%%%%%%%%%%%%%%%%%%%%%%%%%%%%%%%%%%%%%%%%%%%%%%%%%%%%%%%%%%%%%%%%%%

\begin{figure}
\centerline{\epsfxsize 4.5 truein \epsfbox {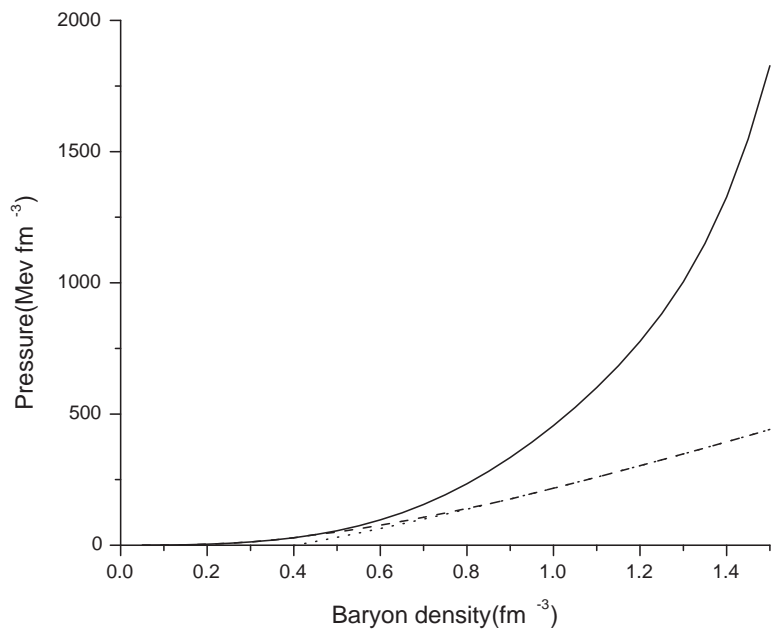}}
\caption{The pressure versus baryon density for Hadron Phase
(solid line), Mixed Phase (dashed line) and Quark Phase (dotted
line) for B=90.} \label{P(rho)90}
\end{figure}

%%%%%%%%%%%%%%%%%%%%%%%%%%%%%%%%%%%%%%%%%%%%%%%%%%%%%%%%%%%%%%%%%%%%%%%%%%%%%%%%

\begin{figure}
\centerline{\epsfxsize 4.5 truein \epsfbox {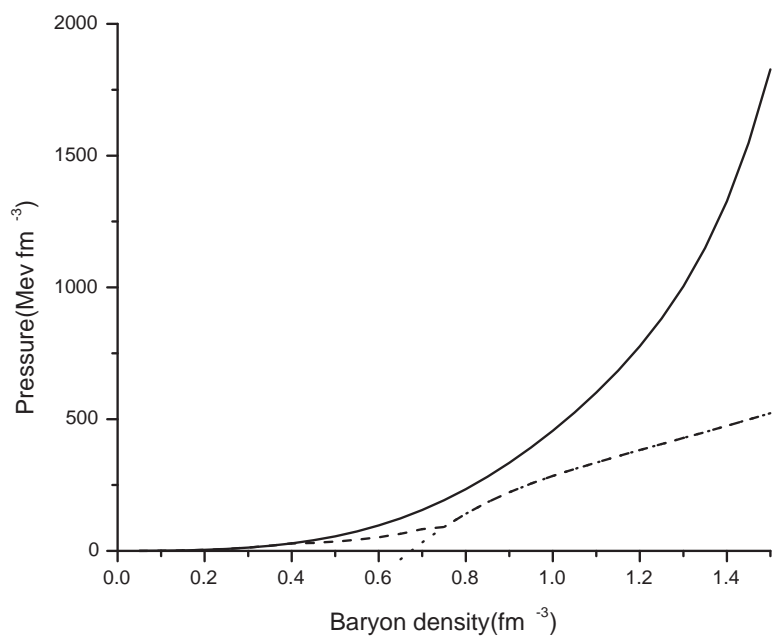}}
\caption{The pressure versus baryon density for Hadron Phase
(solid line), Mixed Phase (dashed line) and Quark Phase (dotted
line) for density dependent B.} \label{P(rho)B(rho)}
\end{figure}

%%%%%%%%%%%%%%%%%%%%%%%%%%%%%%%%%%%%%%%%%%%%%%%%%%%%%%%%%%%%%%%%%%%%%%%%%%%%%%

\begin{figure}
\centerline{\epsfxsize 4.5 truein \epsfbox {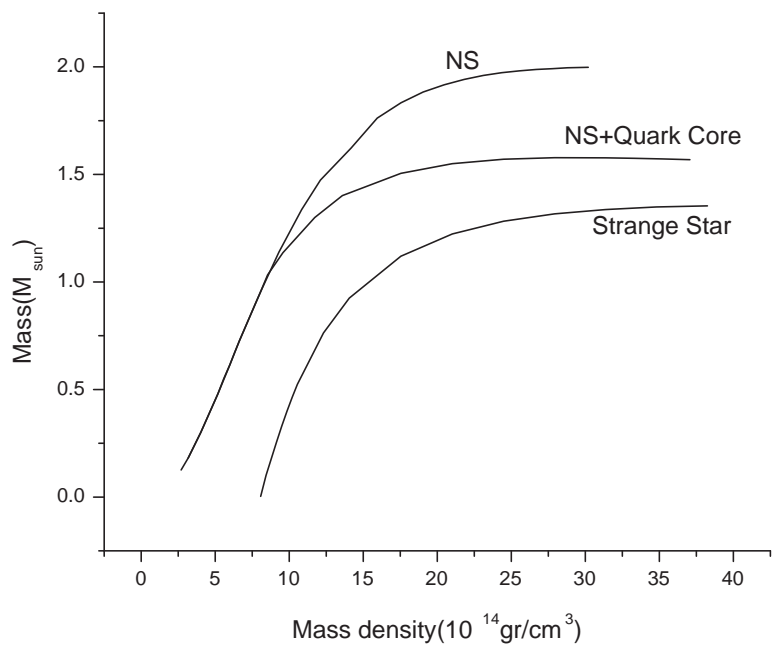}}
\caption{The gravitational mass versus central mass density for
different cases with bag constant B=90.} \label{meB90}
\end{figure}

%%%%%%%%%%%%%%%%%%%%%%%%%%%%%%%%%%%%%%%%%%%%%%%%%%%%%%%%%%%%%%%%%%%%%%%%%%%%%%%%

\begin{figure}
\centerline{\epsfxsize 4.5 truein \epsfbox {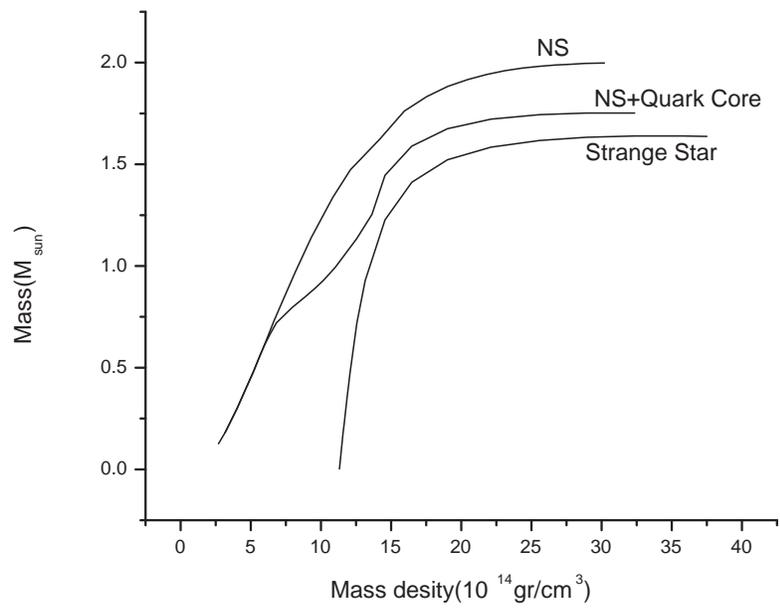}}
\caption{The gravitational mass versus central mass density for
different cases with density dependent B.} \label{meBp}
\end{figure}

%%%%%%%%%%%%%%%%%%%%%%%%%%%%%%%%%%%%%%%%%%%%%%%%%%%%%%%%%%%%%%%%%%%%%%%%%%%%%%%%

\begin{figure}
\centerline{\epsfxsize 4.5 truein \epsfbox {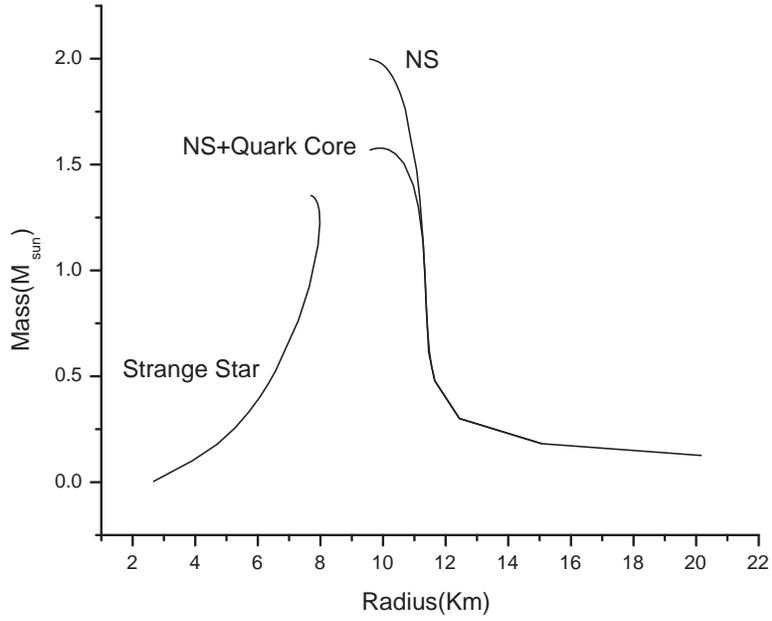}}
\caption{The mass-radius relation for different cases with bag
constant B=90.} \label{mRB90}
\end{figure}

%%%%%%%%%%%%%%%%%%%%%%%%%%%%%%%%%%%%%%%%%%%%%%%%%%%%%%%%%%%%%%%%%%%%%%%%%%%%%%

\begin{figure}
\centerline{\epsfxsize 4.5 truein \epsfbox {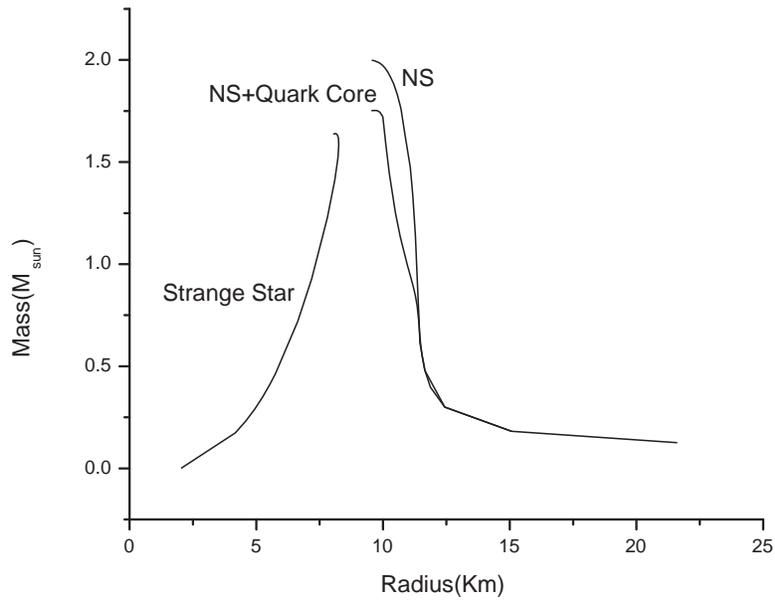}}
\caption{The mass-radius relation for different cases with density
dependent B.}\label{mRBp}
\end{figure}

%%%%%%%%%%%%%%%%%%%%%%%%%%%%%%%%%%%%%%%%%%%%%%%%%%%%%%%%%%%%%%%%%%%%%%%%%%%%%%%%%

\begin{figure}
\centerline{\epsfxsize 4.5 truein \epsfbox {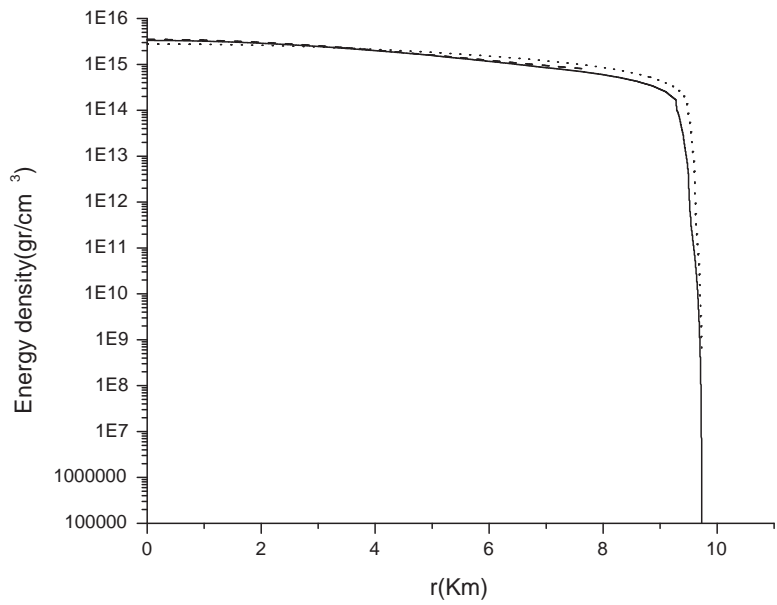}}
\caption{Mass density as a function of radial coordinate for
neutron star (dotted line) neutron star with quark core (solid
curve) and strange star (dashed curve) with B=90.} \label{erB90}
\end{figure}

%%%%%%%%%%%%%%%%%%%%%%%%%%%%%%%%%%%%%%%%%%%%%%%%%%%%%%%%%%%%%%%%%%%%%%%%%%%%%%%%%%

\begin{figure}
\centerline{\epsfxsize 4.5 truein \epsfbox {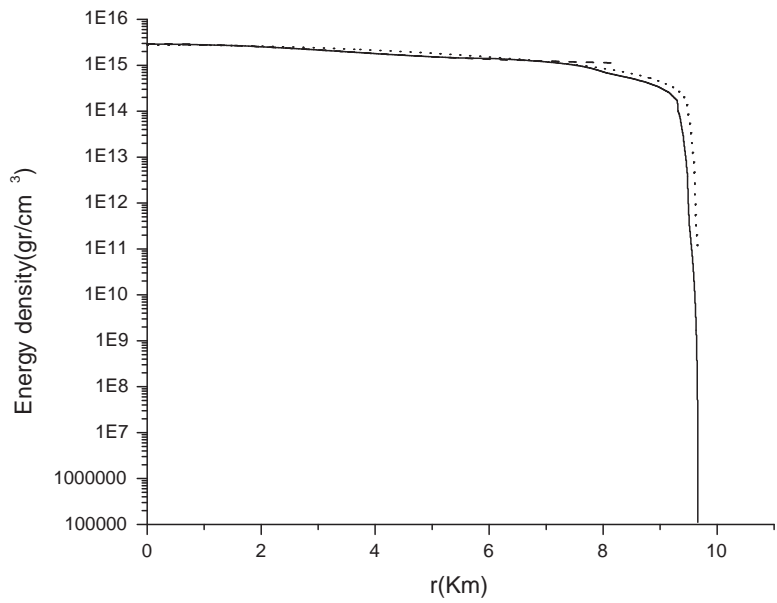}}
\caption{Mass density as a function of radial coordinate for
neutron star (dotted line) neutron star with quark core (solid
curve) and strange star (dashed curve)with density dependent B.}
\label{erBp}
\end{figure}

\end{document}